\documentclass[prb]{revtex4}

\usepackage{graphicx}
\usepackage{dcolumn}
\usepackage{bm}

\begin{document}


\title{Tunneling-induced high efficiency four-wave mixing in an asymmetric quantum wells}

\author{Hui Sun$^{1}$}
\author{Shuangli Fan$^{1}$}\thanks{fanshuangli@snnu.edu.cn}
\author{Hongjun Zhang$^{1}$}
\author{Shangqing Gong$^{2}$}\thanks{sqgong@ecust.edu.cn}

\affiliation{$^{1}$School of Physics and Information Technology,
Shaanxi
Normal University, Xi'an 710062, China\\
$^{2}$Department of Physics, East China University of Science and
Technology, Shanghai 200237, China}

\date{\today}

\begin{abstract}
An asymmetric double quantum wells (QWs) structure with resonant
tunneling is suggested to achieve high efficient four wave mixing
(FWM). We analytically demonstrate that the resonant tunneling can
induce high efficient mixing wave in such a semiconductor structure
with a low light pump wave. In particular, the FWM conversion
efficiency can be enhanced dramatically in the vicinity of the
center frequency. This interesting scheme may be used to generate
coherent long-wavelength radiation in solid-state system.
\end{abstract}

\maketitle


Multiwave mixing processes in the ultraslow propagating regime has
been an area of active research for many years. This mainly due to
the potentially wide range of applications in diverse fields as well
as high-efficiency generation of coherent radiation, optical image
amplification, quantum nonlinear optics, quantum information
science, and so on. As one of the centerpieces of modern technology,
how to realize efficiency generation of coherent light has been of
primary importance. Harris and
co-workers~\cite{harris-prl-1990,harris-prl-1999} have introduced
the idea of electromagnetically induced transparency
(EIT)~\cite{fleischhauer-rmp-2005}, which allows controlled
manipulations of the optical properties of atom or atom-like media
and hence leads to suppression of linear absorption and enhancement
of nonlinear susceptibility. This means that EIT-based high
efficient four-wave mixing (FWM) process is hopefully possible to
obtain by eliminating linear absorption. The FWM in EIT media has
been the focus of several recent
studies~\cite{hemmer-ol-1995,deng-prl-2002,deng-pra-2003,li-pra-2007,yang-jpb-2005,
niu-pra-2005,wu-ol-2004,wu-pra-2004}. Hemmer and coworkers
introduced the double $\Lambda$-scheme as an important tool in
EIT-based resonant FWM~\cite{hemmer-ol-1995}. Deng
\textit{et.al.}~\cite{deng-prl-2002,deng-pra-2003} suggested a novel
channel opening technique of FWM in a four-level system and showed
that nonexisting wave-mixing channels may be opened deeply inside
the medium and the efficiency of the FWM can be significantly
enhance. Subsequently, the efficient multiwave mixing resulted from
multiphoton destructive interference~\cite{wu-pra-2004} and
double-dark resonances~\cite{niu-pra-2005,yang-jpb-2005,li-pra-2007}
have been proposed. Experimentally, the results on FWM in a thermal
vapor cell of rubidium atoms via highly excited Rydberg states has
been presented~\cite{kolle-pra-2012}.

FWM in atomic vapors has been explored extensively in the past. Much
of the recent literature has exploited in semiconductor quantum
wells (QWs) structure. QWs structure provides a potential energy
well, the confined two dimension electron gas behaves atomiclike
optical responses. Different from atomic system, the advantages of
semiconductor QWs structure such as large electric dipole momentums,
controllable intersubband energies and the electron function
symmetries creat the opportunities of building opto-electron devices
that harness atom physics. There is great interest in extending the
study of quantum interference phenomena from atomic system to
semiconductor QWs structure, and several quantum optical coherence
and interference
effects~\cite{phillips-prl-2003,joshi-prb-2009,wu-prl-2005,wu-pra-2006,
yuan-apl-2006,sun-prb-2006,sun-prb-2009-1,sun-prb-2009-2,
paspalakis-prb-2004,paspalakis-prb-2006-1,kosionis-jap-2011,
paspalakis-prb-2006-2,zhu-oe-2011,zhu-prb-2009,yang-pra-2008,
ginzburg-oe-2006,sun-ol-2007,sun-oe-2012} have been studied
theoretically and experimentally. For examples,
EIT~\cite{phillips-prl-2003,joshi-prb-2009}, ultrafast all-optical
switching via Fano interference~\cite{wu-prl-2005}, slow
light~\cite{yuan-apl-2006,ginzburg-oe-2006} and optical solitons via
interband transitions~\cite{zhu-prb-2009}, tunneling-induced
enhancement of Kerr nonlinearity~\cite{sun-ol-2007,sun-oe-2012}, and
so on. FWM processes in semiconductor QWs structure have also been
paid much attention. The idea of resonant FWM in double
$\Lambda$-scheme~\cite{hemmer-ol-1995} has been extended into
asymmetric double QWs structure based on
interband~\cite{hao-pla-2008} and intersubband
transitions~\cite{yang-jmo-2009}, respectively. Subsequently, a
cascade configuration for obtaining highly efficient FWM process in
three coupled QWs based on intersubband transition was
suggested~\cite{hao-pla-2009}. More recently, FWM in the diamond
configuration in an atomic vapor~\cite{wills-pra-2009} and
X-ray/optical sum-frequency generation in
diamond~\cite{glover-nature-2012} have been demonstrated
experimentally. FWM process in quantum dot (QD) nanostructure has
been suggested theoretically~\cite{hao-josab-2012}.

\begin{figure}
\includegraphics[width=6.5cm]{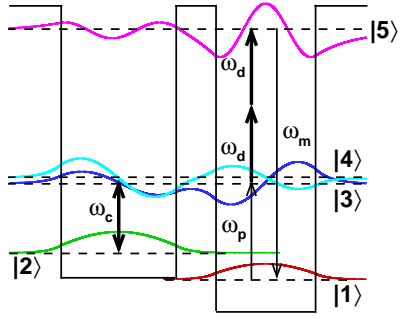}
\caption{(color online) Conduction subband of the asymmetric double
quantum wells (QWs) structure, in which two ground subbands
$|1\rangle$ and $|2\rangle$ are, respectively, driven by a weak pulsed pump
and a continuous-wave control fields via the subbands $|3\rangle$ and
$|4\rangle$. The subbands $|3\rangle$ and $|3\rangle$ are coupled with
$|5\rangle$ by a two-photon process. The solid curves represent the
corresponding wave functions.}\label{fig:band-structure}
\end{figure}

In QWs structure, the resonant tunneling induces not only
transparency, but also the enhancement of Kerr
nonlinearity~\cite{sun-ol-2007,sun-oe-2012}. Thus, the natural
question is can we enhance the efficiency of FWM in QWs structure
via resonant tunneling. In the present paper, we suggest an
asymmetric double QWs structure and address this question. The basic
idea is to combine the constructive interference in Kerr
nonlinearity and EIT-based channel opening technique. Our studies
demonstrate that with two continuous-wave (cw) laser fields and a
weak, pulsed pump field, a pulsed FWM field can be generated. With
the presence of resonant tunneling, the enhancement of conversion
efficiency of FWM can be achieved. In particular, around the center
frequency, the FWM conversion efficiency can be enhanced
dramatically. This QWs structure may be useful to generate coherent
long wavelength radiation in solid materials.


The asymmetric double QWs structure, which interacts with three
laser fields, is shown in Fig.~\ref{fig:band-structure}. This
structure consists of a Al$_{0.04}$Ga$_{0.96}$As layer (shallow
well) with thickness of 11.0~nm and a 9.5~nm GaAs layer (deep well),
and they are separated by a 3.8~nm Al$_{0.4}$Ga$_{0.6}$As potential
barrier. Both the left side of shallow well and the right side of
deep well are Al$_{0.4}$Ga$_{0.6}$As potential barriers. In this
structure, $|1\rangle$ and $|2\rangle$ with energies $34.5$~meV and
$62.3$~meV are, respectively, the ground subbands of the left
shallow and the right deep wells. Two subbands $|3\rangle$ and
$|4\rangle$ with eigenenergies 135.5~meV and 141.5~meV are created
by mixing the first excited subbands of the shallow ($|se\rangle$)
and deep ($|de\rangle$) wells by resonant tunneling. Their
corresponding wave functions are symmetric and asymmetric
combinations of $|se\rangle$ and $|de\rangle$, i.e.,
$|3\rangle=(|se\rangle-|de\rangle)/\sqrt{2}$ and
$|4\rangle=(|se\rangle+|de\rangle)/\sqrt{2}$. $|5\rangle$ is the
second excited subband of the right deep well, and its energy is
296.3~meV. A weak pulsed pump ($\omega_{p}$) and a continuous wave
($\omega_{c}$) fields, respectively, couple the ground subbands
$|1\rangle$ and $|2\rangle$ via two short-lived subbands $|3\rangle$
and $|4\rangle$. Subbands $|1\rangle$, $|3\rangle$, and $|4\rangle$
form the usual TIT scheme together with the pump field. The driven
field ($\omega_{d}$) provides a two-photon coupling to the FWM
subband $|5\rangle$, which is dipole coupled to the ground subband
$|1\rangle$. The pump field and the resonant two-photon field induce
the FWM processes $|1\rangle\to|3\rangle\to|5\rangle\to|1\rangle$
and $|1\rangle\to|4\rangle\to|5\rangle\to|1\rangle$, then generate a
coherent radiation field at frequency $\omega_{m}$ with
$\omega_{m}=\omega_{p}+2\omega_{d}$. The solid curves represent the
corresponding wave functions. Without the subband $|4\rangle$, the
scheme is simplified as that in Ref.~\cite{deng-prl-2002}. The
novelty of this scheme under consideration is that it combines the
advantages of the Deng-scheme~\cite{deng-prl-2002} and constructive
interference associated with Kerr nonlinearity induced by resonant
tunneling~\cite{sun-ol-2007,sun-oe-2012}. The band structure is
designed with large electronic dipole momentum $\mu_{15}$, which can
enhance the FWM efficiency.

Under the dipole and rotating-wave approximations (RWA), the
structure dynamics can be described using equations of motion for
probability amplitudes $b_{i}(t)$ $(i=1-5)$ of the states
\begin{eqnarray}
&&i\partial_{t}b_{1}+\Omega_{p}b_{3}+m\Omega_{p}b_{4}+\Omega_{m}b_{5}=0,\label{eq-1-sch-eq-b1}\\
&&(i\partial_{t}+d_{2})b_{2}+\Omega_{c}b_{3}+q\Omega_{c}b_{4}=0,\label{eq-1-sch-eq-b2}\\
&&(i\partial_{t}+d_{3})b_{3}+\Omega_{p}b_{1}+\Omega_{c}b_{2}+\Omega_{d}^{(2)}e^{-i\Delta
kz}b_{5}=0,\label{eq-1-sch-eq-b3}\\
&&(i\partial_{t}+d_{4})b_{4}+m\Omega_{p}b_{1}+q\Omega_{c}b_{2}
+k\Omega_{d}^{(2)}e^{-i\Delta
kz}b_{5}=0,\label{eq-1-sch-eq-b4}\\
&&(i\partial_{t}+d_{5})b_{5}+\Omega_{m}b_{1}+\Omega_{d}^{(2)}e^{i\Delta
kz}(b_{2}+kb_{5})=0.\label{eq-1-sch-eq-b5}
\end{eqnarray}
As usual, $\Omega_{p}=\mu_{31}E_{p}/2\hbar$,
$\Omega_{c}=\mu_{32}E_{c}/2\hbar$, and
$\Omega_{m}=\mu_{51}E_{m}/2\hbar$ are, one half of the Rabi
frequencies for the respectively transitions with $\mu_{31}$,
$\mu_{32}$, $\mu_{51}$ being corresponding dipole matrix elements,
and $\Omega_{d}^{(2)}$ is the direct two-photon Rabi frequency from
subband $|3\rangle$ to $|5\rangle$. $m=\mu_{41}/\mu_{31}$,
$q=\mu_{42}/\mu_{32}$, and $k=\mu_{54}/\mu_{53}$ give the ratios
between the relevant subband transition dipole momentum. In the QWs
structure under consideration, it can be calculated that $m=-0.42$,
$q=1.63$, and $k=-1.12$. $d_{2}=\Delta_{p}-\Delta_{c}+i\gamma_{2}$,
$d_{3}=\Delta_{p}+i\gamma_{3}$,
$d_{4}=\Delta_{p}-\delta+i\gamma_{4}$, and
$d_{5}=\Delta_{m}+i\gamma_{5}$ with
$\Delta_{p(c,m)}=\omega_{p(c,m)}-(\omega_{3(3,5)}-\omega_{1(2,1)})$
being detunings, and $\gamma_{j}$ ($j=2$ to 5) being the electron
decay rate of the state $|j\rangle$. In a usual way, $\Delta
k=k_{p}+2k_{d}-k_{m}$ denotes phase mismatching. In order to
correctly predict the efficiency of FWM field generated, equations
(\ref{eq-1-sch-eq-b1})-(\ref{eq-1-sch-eq-b5}) must be solved
together with the wave equations for the pump and FWM fields. Under
the slowly varying envelop approximation, they read
\begin{eqnarray}
&&\frac{\partial\Omega_{p}(z,t)}{\partial
z}+\frac{1}{c}\frac{\partial\Omega_{p}(z,t)}{\partial
t}=i\kappa_{p}(b_{3}+mb_{4})b_{1}^{*},\label{eq-1-wave-eq-p}\\
&&\frac{\partial\Omega_{m}(z,t)}{\partial
z}+\frac{1}{c}\frac{\partial\Omega_{m}(z,t)}{\partial
t}=i\kappa_{m}b_{5}b_{1}^{*}.\label{eq-1-wave-eq-m}
\end{eqnarray}
Here, the propagation constant
$\kappa_{p(m)}=N\omega_{p(m)}|\mu_{31(51)}|^{2}/(\hbar\varepsilon_{0}c)$,
with $N$ being the electron concentration, and $c$ the speed of
light in vacuum. In the following, we assume $\Omega_{p}$ and the
two-photon transition field $\Omega_{d}^{(2)}$ are very
comparatively weak (i.e., considering the FWM process at low pump
intensity), the electronic ground subband $|1\rangle$ is not
depleted, i.e., $b_{1}\approx1$. Under these assumptions, following
the standard processes~\cite{deng-prl-2002}, one immediately has, in
dimensionless form,
\begin{eqnarray}
&&\alpha_{3}+m\alpha_{4}=-\frac{W_{p}\tau D_{p}}{D},\label{eq-1-alpha34}\\
&&\alpha_{5}=-\frac{W_{m}\tau}{\eta+d_{5}\tau}+\frac{W_{p}\tau\Omega_{d}^{(2)}\tau
e^{i\Delta kz}D_{m}}{(\eta+d_{5}\tau)D},\label{eq-1-alpha5}
\end{eqnarray}
in which $\alpha_{j}$ ($j=3-5$), $W_{p}$ and $W_{m}$ are the Fourier
transforms of $b_{j}$, $\Omega_{p}$, and $\Omega_{m}$.
$\eta=\omega\tau$ is the dimensionless Fourier transform variable.
$D$, $D_{p}$, and $D_{m}$ are, respectively, given by
\begin{eqnarray*}
&&D=(\eta+d_{2}\tau)(\eta+d_{3}\tau)(\eta+d_{4}\tau)-[(\eta+d_{4}\tau)+q^{2}(\eta+d_{3}\tau)](\Omega_{c}\tau)^{2},\\
&&D_{p}=(\eta+d_{2}\tau)[(\eta+d_{4}\tau)+m^{2}(\eta+d_{3}\tau)]-(q-m)^{2}(\Omega_{c}\tau)^{2},\\
&&D_{m}=(\eta+d_{2}\tau)[(\eta+d_{4}\tau)+mk(\eta+d_{3}\tau)]-(q-k)(q-m)(\Omega_{c}\tau)^{2}.
\end{eqnarray*}
Taking Fourier transform of
Eqs.~(\ref{eq-1-wave-eq-p})-(\ref{eq-1-wave-eq-m}), and combining
the results of (\ref{eq-1-alpha34}) and (\ref{eq-1-alpha5}), we
obtain
\begin{eqnarray}
&&W_{m}(z,\eta)=\frac{\kappa_{m}c\tau^{2}\Omega_{d}^{(2)}\tau
D_{m}W_{p}(0,\eta)}{(\eta+d_{5}\tau)D}\frac{e^{iBz/c\tau}-1}{B}\exp\left\{\frac{i\eta
z}{c\tau}
\left[1-\frac{\kappa_{m}c\tau^{2}}{\eta(\eta+d_{5}\tau)}\right]\right\}.\label{eq-1-wm}
\end{eqnarray}
Here, $W_{p}(0,\eta)$ is the Fourier transform of $\Omega_{p}(z,t)$
at the entrance of the QWs structure, and $B$ is given by
\begin{eqnarray*}B=\Delta
k\tau+\kappa_{m}c\tau^{2}/(\eta+d_{5}\tau)-\kappa_{p}c\tau^{2}D_{p}/D.
\end{eqnarray*}
For the FWM emission process, the boundary condition
$W_{m}(0,\eta)=0$ is applied. Equation (\ref{eq-1-wm}) is the main
result of the present paper, it gives the generated FWM field
$W_{m}(z,\eta)$ at arbitrary frequency detunings. The expression of
$D_{p(m)}$ exhibits the role of resonant tunneling clearly. In the
QWs structure suggested, resonant tunneling leads to the asymmetric
and symmetric wave functions of subbands $|3\rangle$ and
$|4\rangle$. As a result, we have $q\cdot m<0$ and $q\cdot k<0$,
which indicate that resonant tunneling can modify the optical
nonlinearity such as enhancement of FWM conversion efficiency [see
Eq.~(\ref{eq-1-wm}) together with $D_{p}$ and $D_{m}$]. By taking
$q=k=m=0$, the subband $|4\rangle$ is decoupled, the scheme is
therefore reduced to Deng-scheme~\cite{deng-prl-2002}. It can be
easily checked that Eq.~(\ref{eq-1-wm}) reduces to the main result
of Ref.~\cite{deng-prl-2002}. The generated FWM power can be
calculated by taking inversion Fourier transform of
Eq.~(\ref{eq-1-wm}). We assume the pump field at the entrance of the
QWs structure has the form~\cite{deng-prl-2002}
\begin{equation}
W_{p}(0,\eta)=\frac{\tau\Omega_{p}(0,0)}{\sqrt{2}}e^{-\eta^{2}/4},
\end{equation}
and define the conversion efficiency of FWM $\rho$
as~\cite{deng-prl-2002,niu-pra-2005}
\begin{eqnarray}
\rho=\left|\frac{\mu_{31}W_{m}(z,\eta)}{\mu_{51}\tau\Omega_{d}^{(2)}\tau
\Omega_{p}(0,0)/\sqrt{2}}\right|^{2}.
\end{eqnarray}

\begin{figure}
\includegraphics[width=4.2cm]{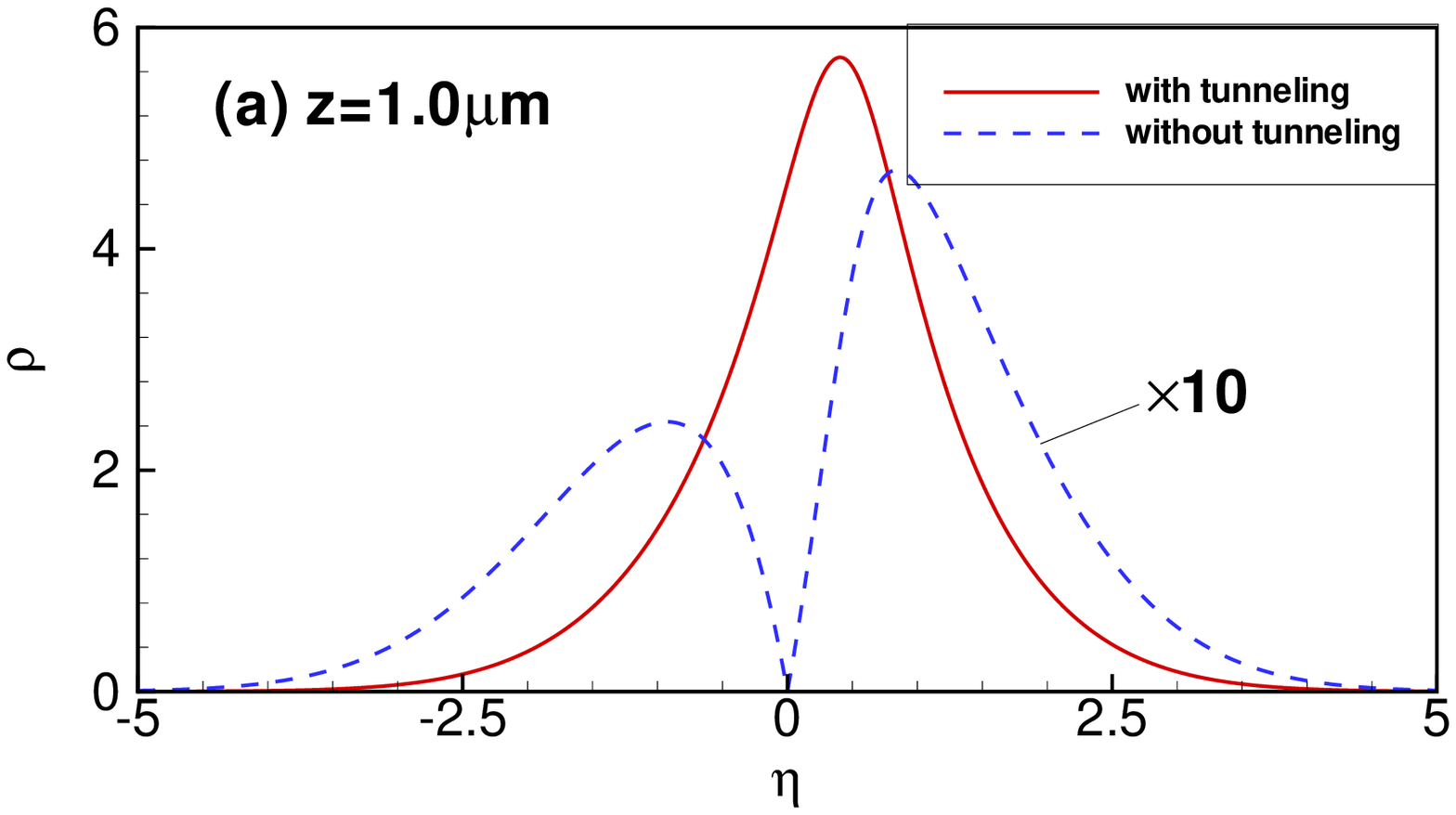}
\includegraphics[width=4.2cm]{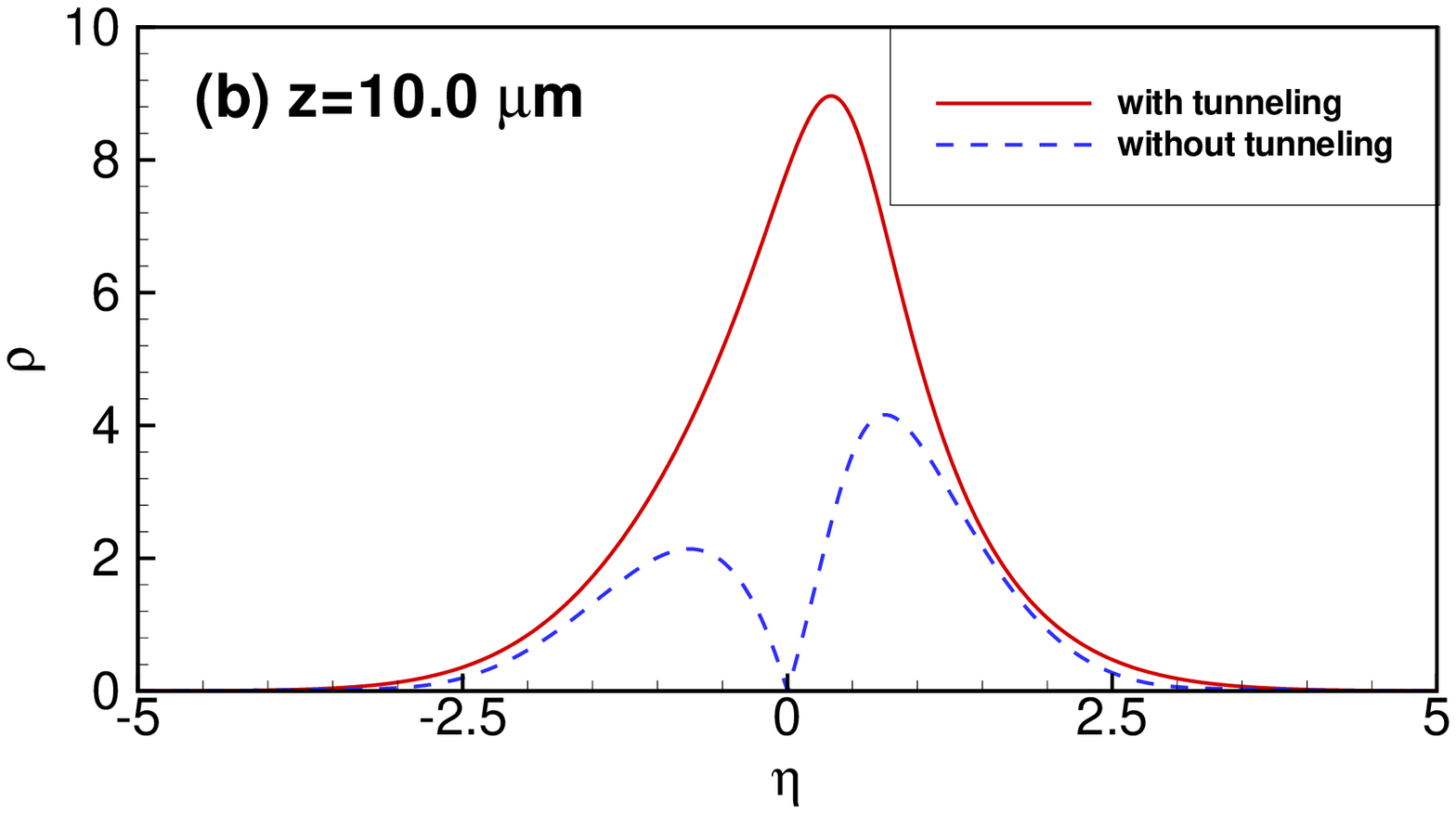}
\includegraphics[width=4.2cm]{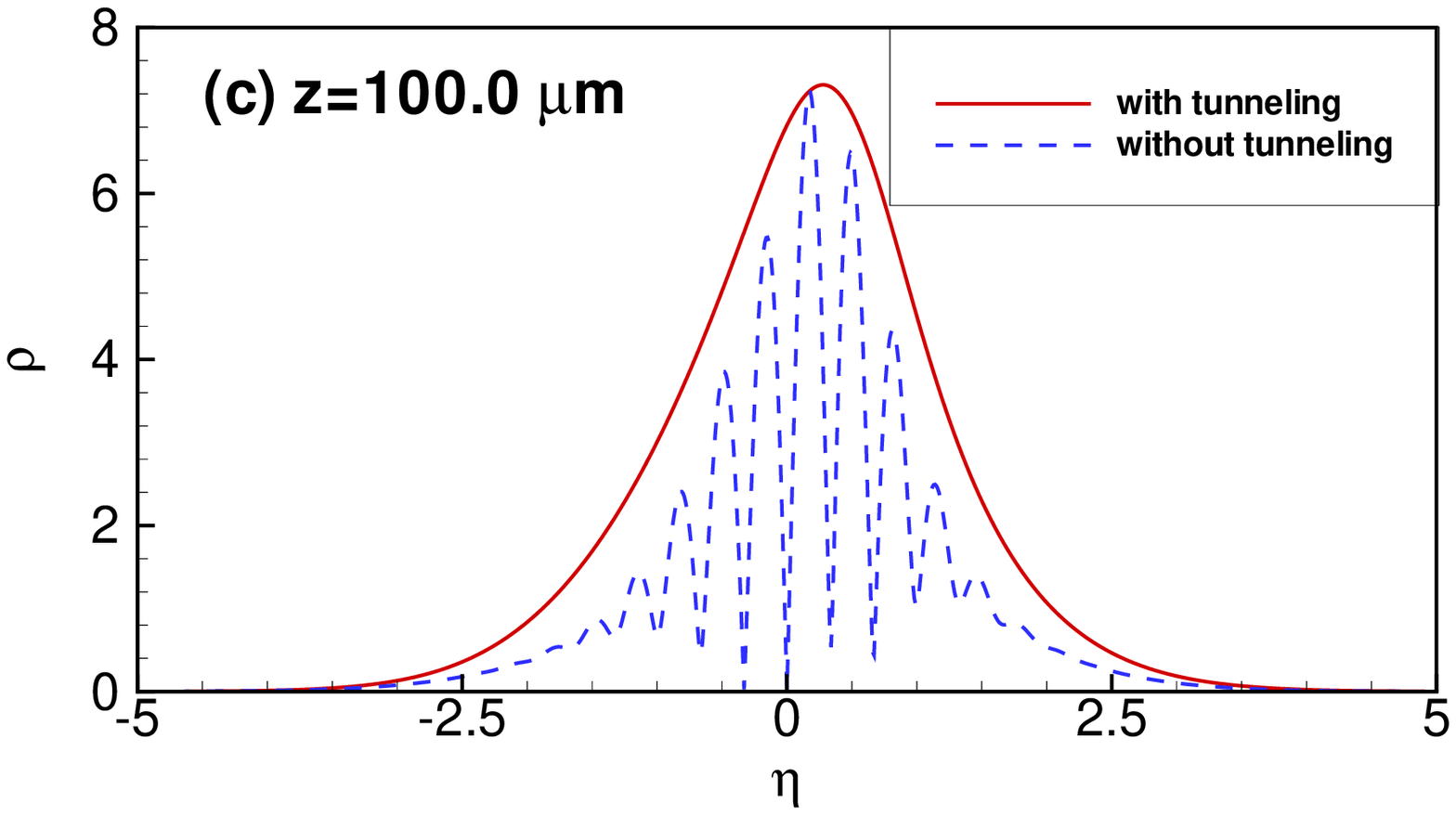}\\
\includegraphics[width=4.2cm]{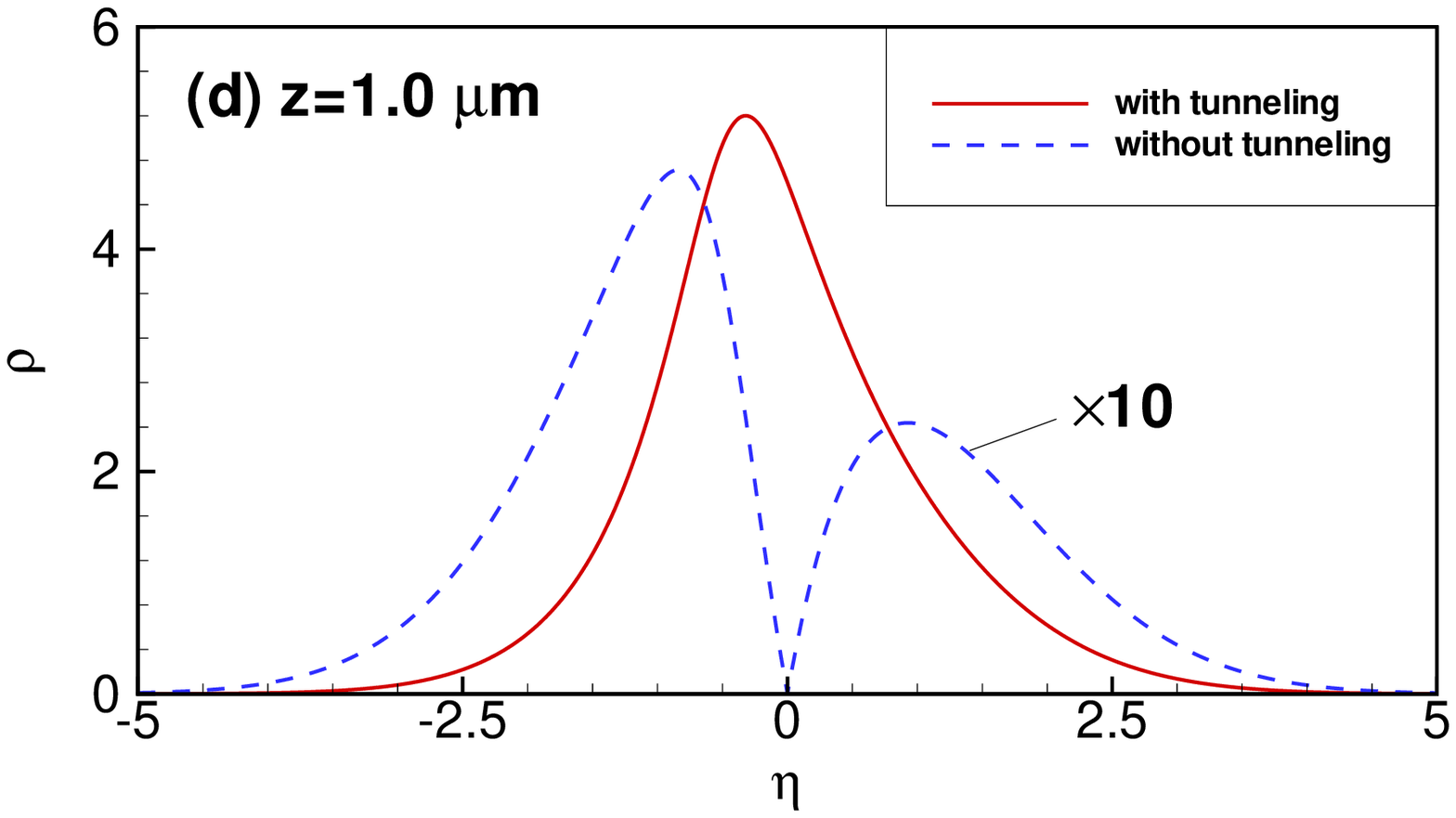}
\includegraphics[width=4.2cm]{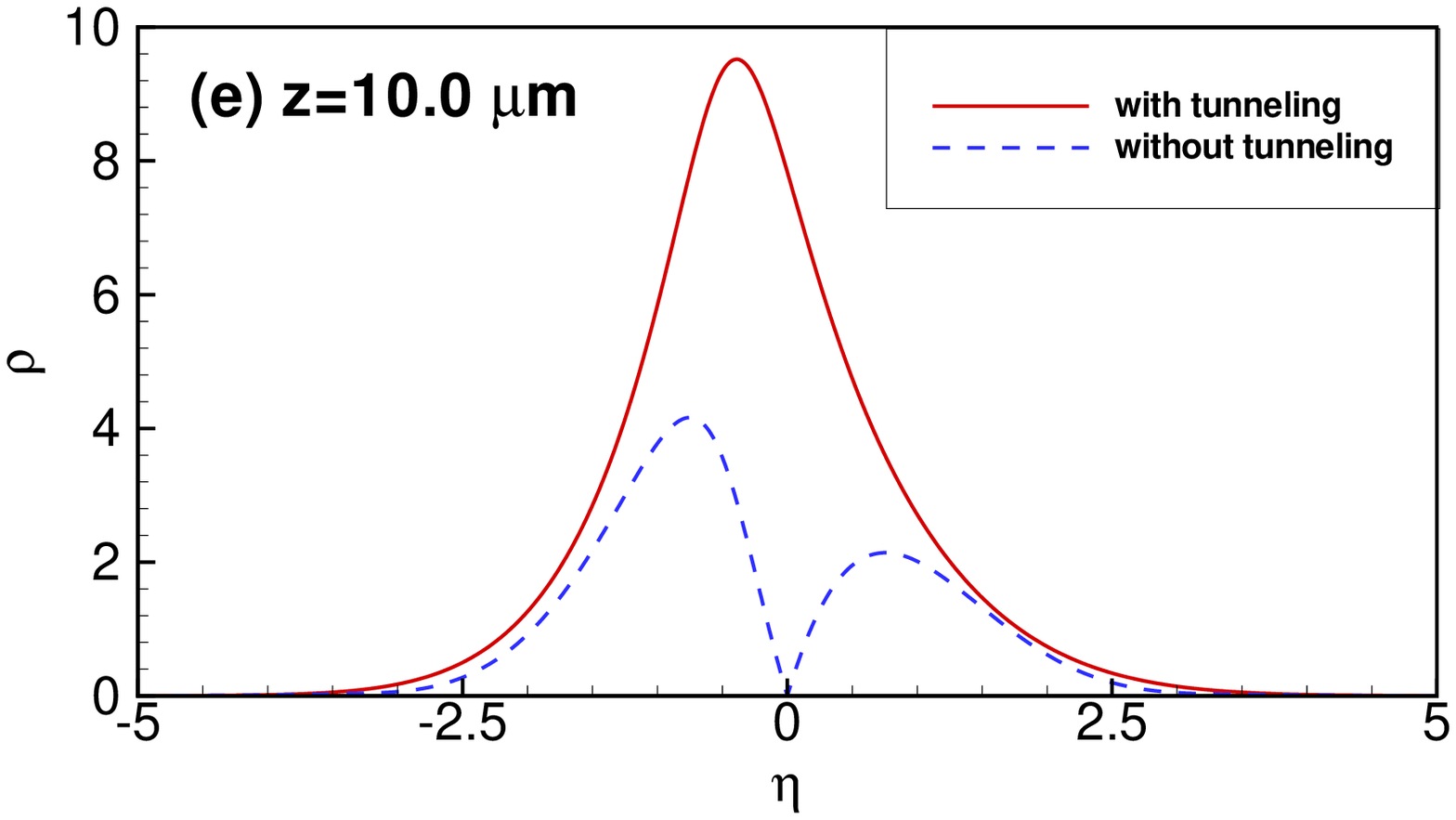}
\includegraphics[width=4.2cm]{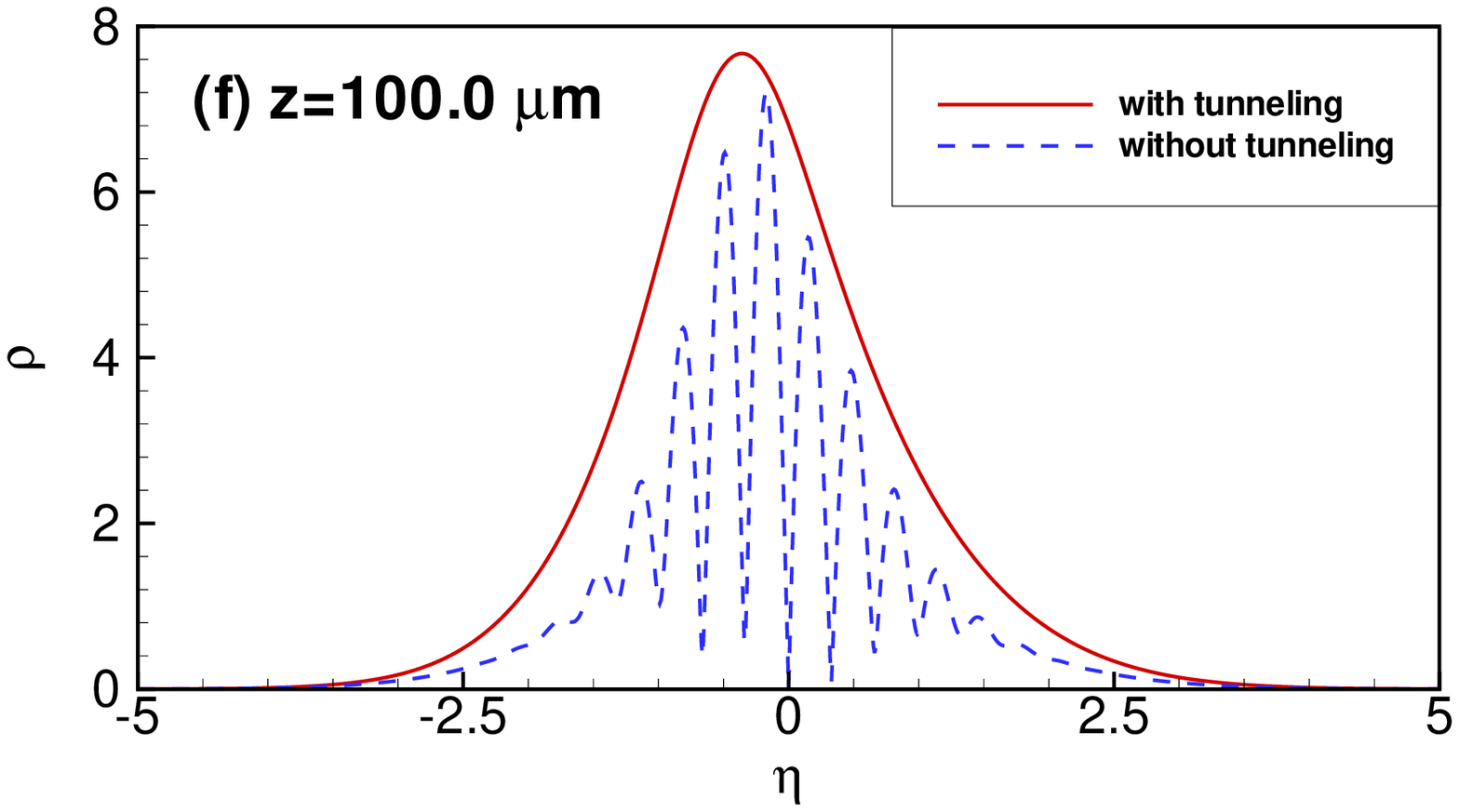}
\caption{(color online) The conversion efficiency of FWM $\rho$ as a
function of $\eta$ with and without resonant tunneling. The values
of $\Delta_{m}$ corresponding to the first and second ranks are,
respectively, $\Delta_{m}=-27.5$~$\mu$eV and
$\Delta_{m}=27.5$~$\mu$eV. The other parameters are explained in the
text.}\label{fig:varrho-eta}
\end{figure}

\begin{figure}
\includegraphics[width=4.2cm]{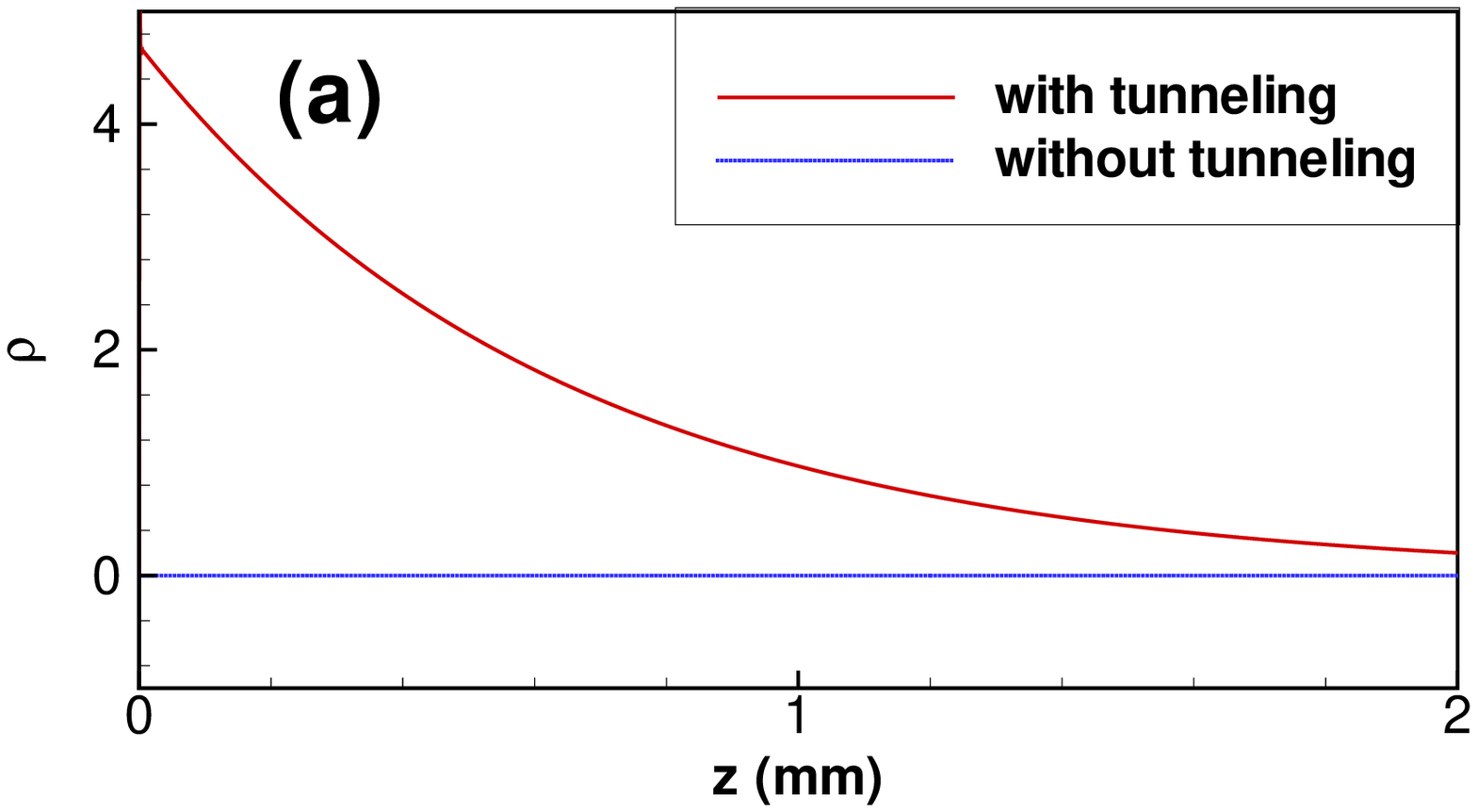}
\includegraphics[width=4.2cm]{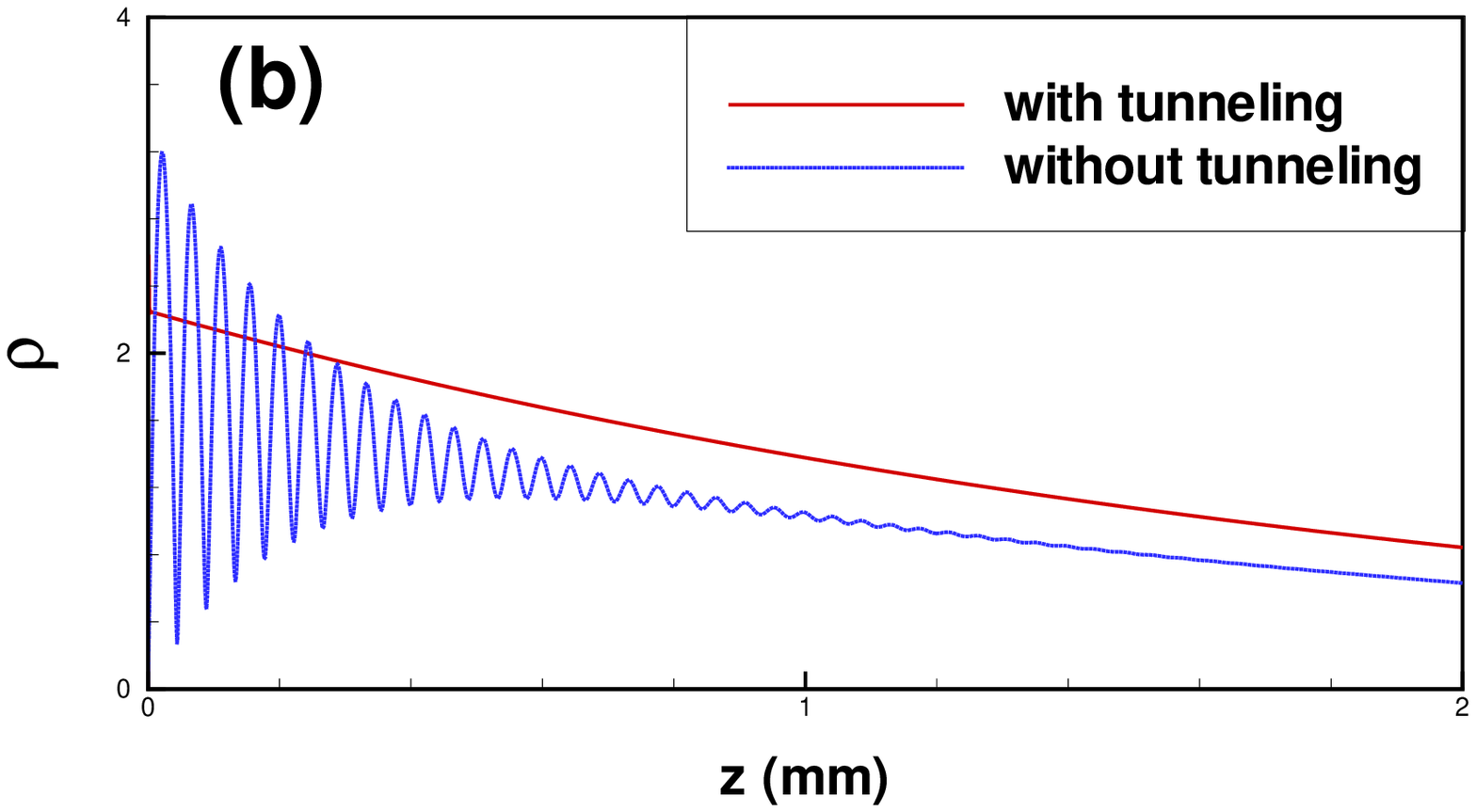}
\includegraphics[width=4.2cm]{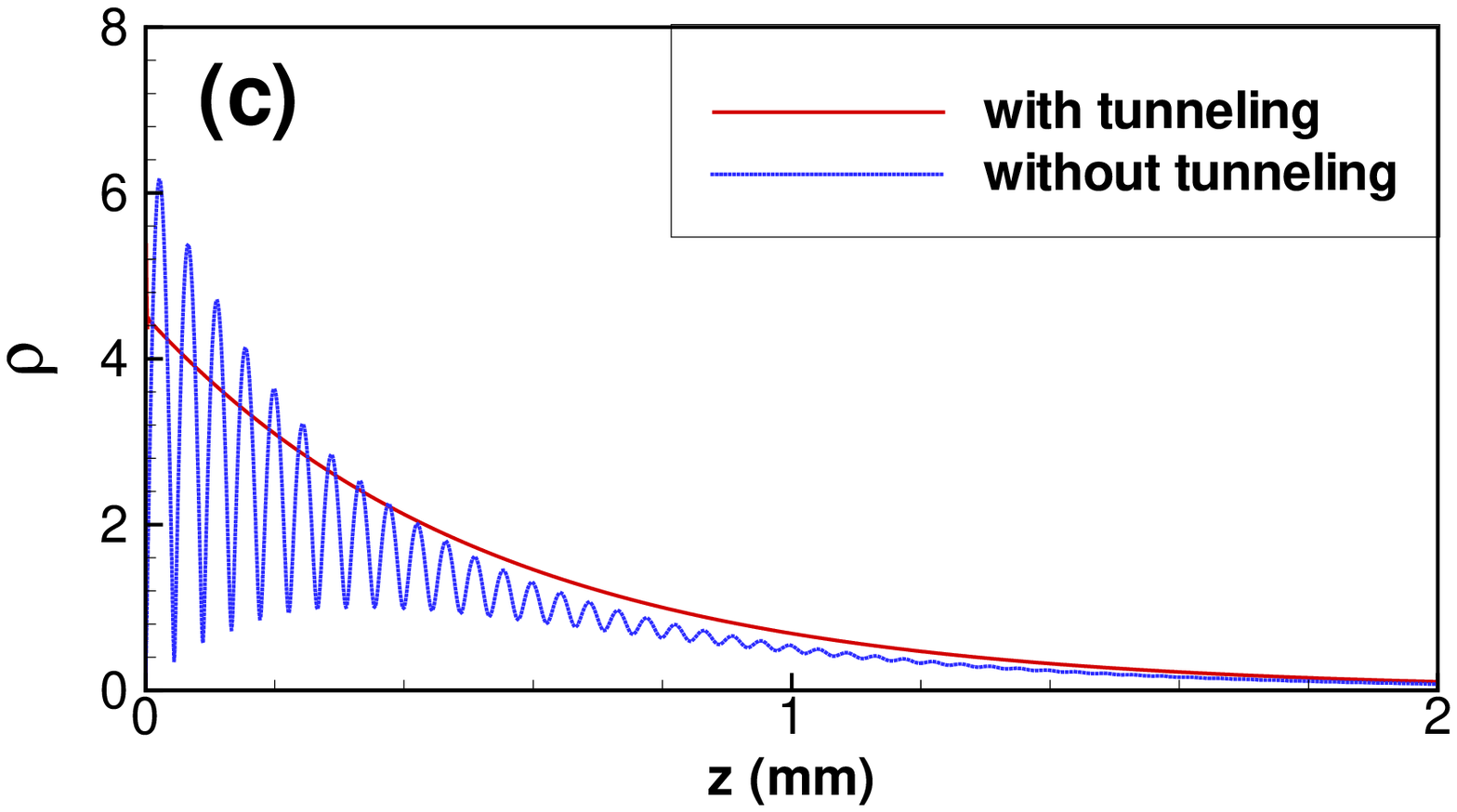}
\includegraphics[width=4.2cm]{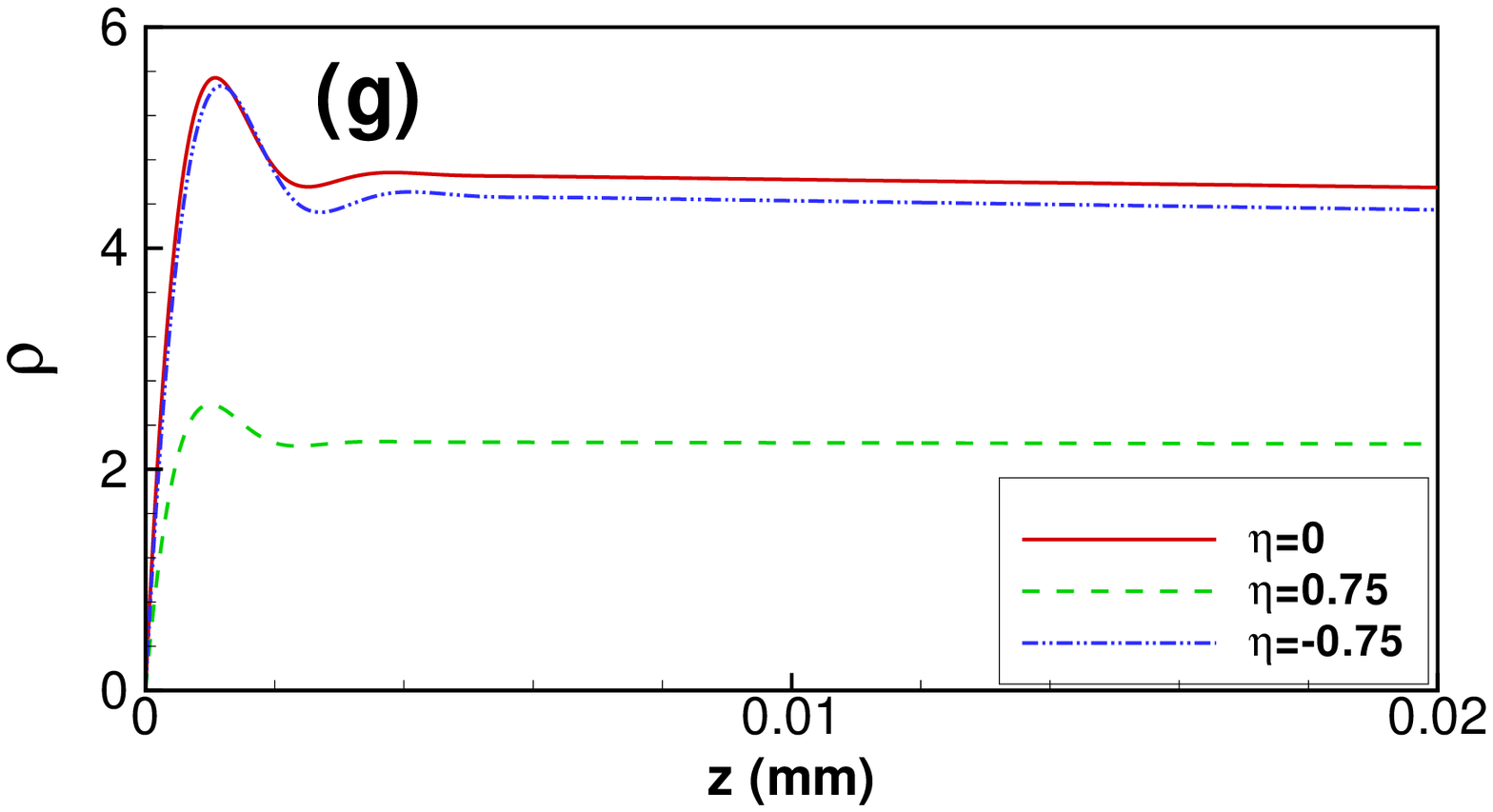}\\
\includegraphics[width=4.2cm]{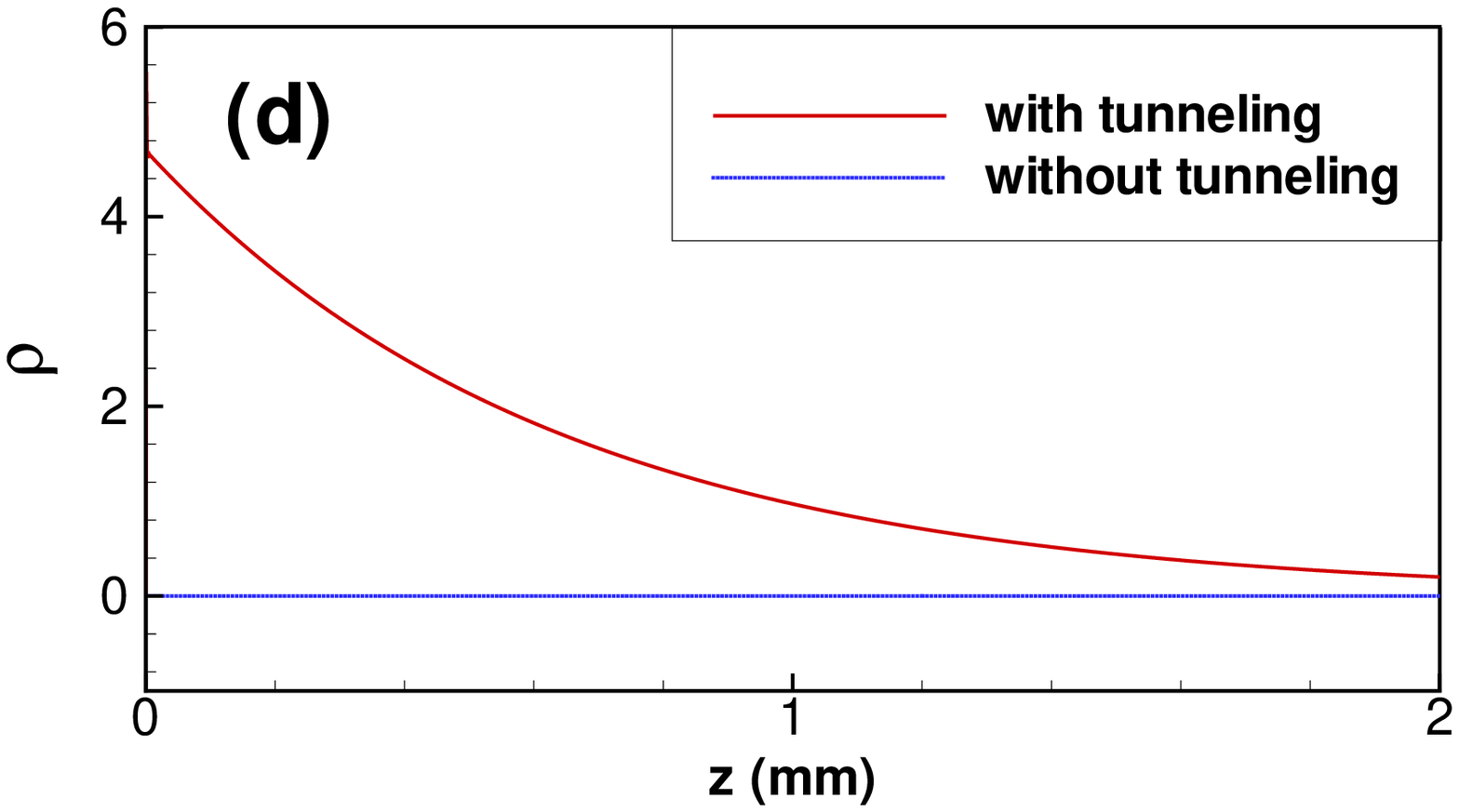}
\includegraphics[width=4.2cm]{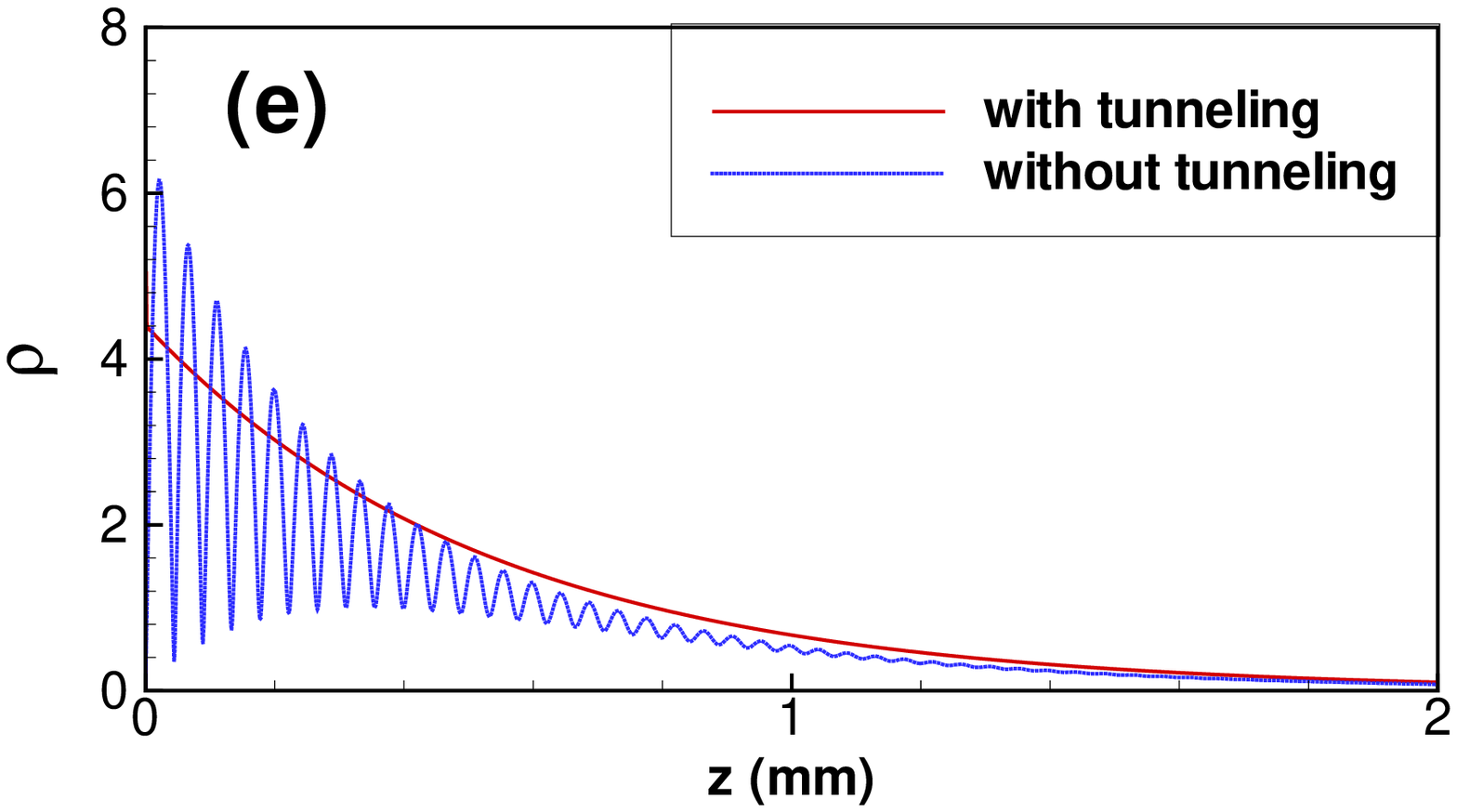}
\includegraphics[width=4.2cm]{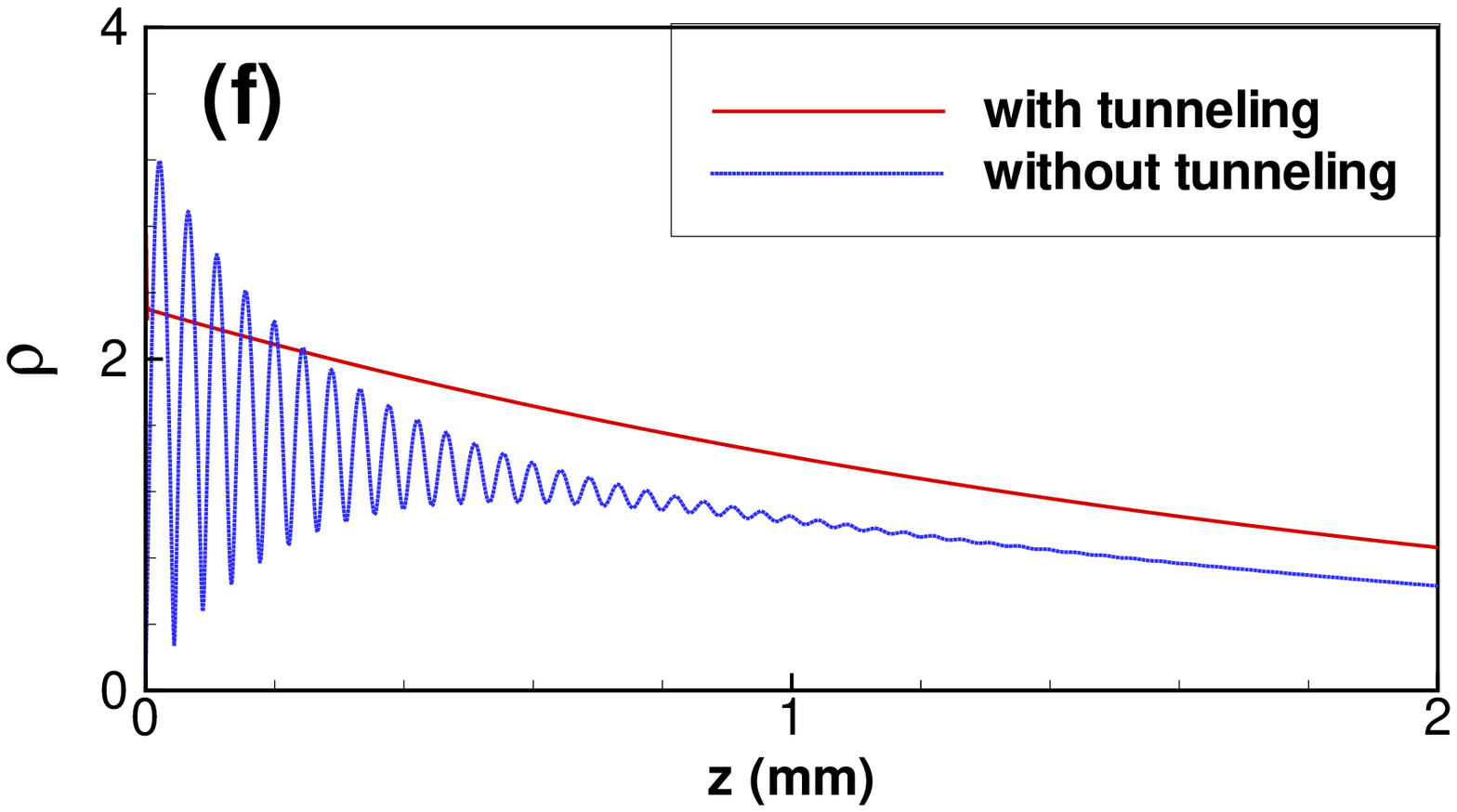}
\includegraphics[width=4.2cm]{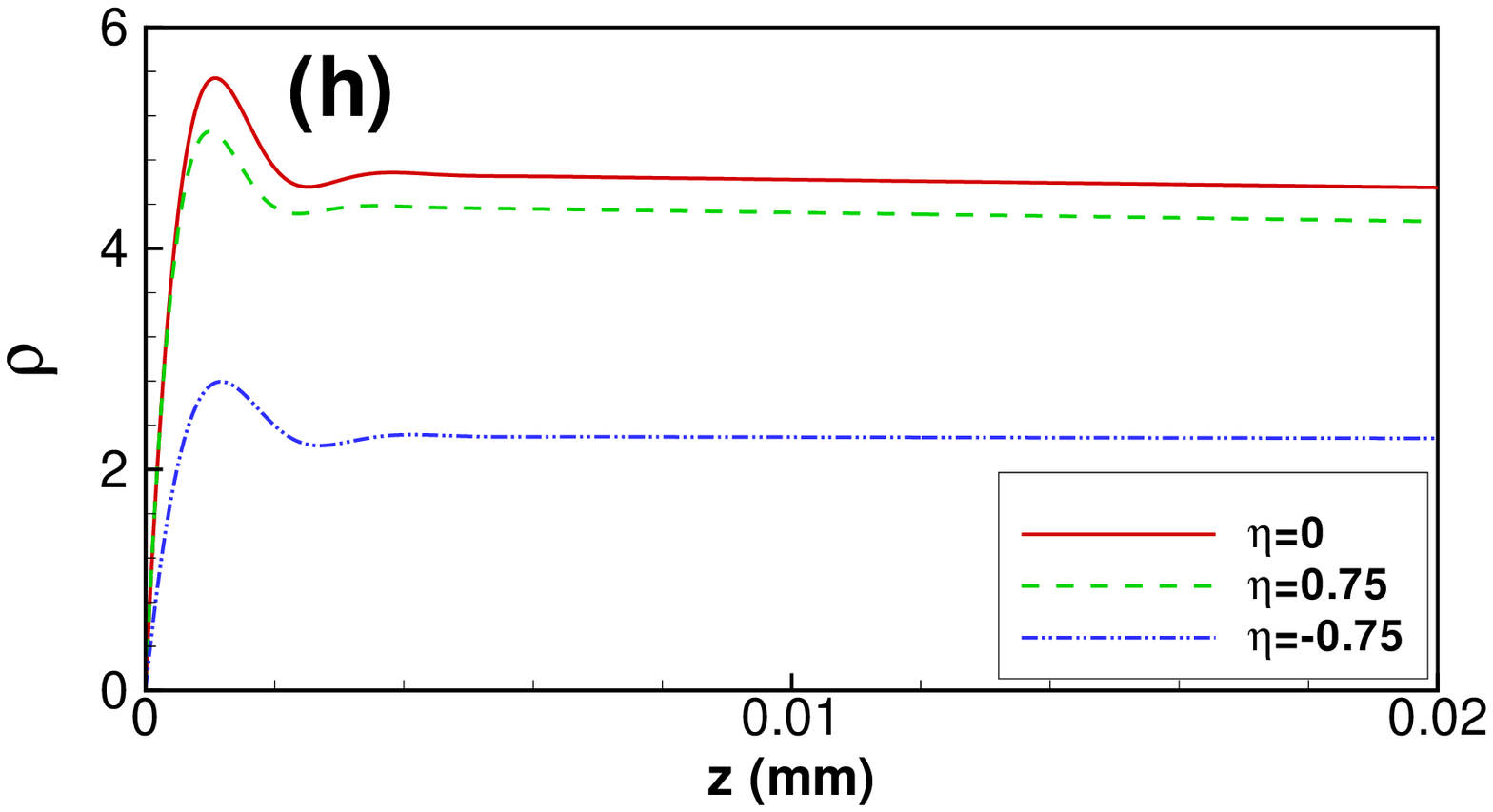}
\caption{(color online) The conversion efficiency of FWM $\rho$ as a
function of propagation distance $z$ with and without resonant
tunneling. The value of $\Delta_{m}$ corresponding to the first and
second ranks are taken as $\Delta_{m}=72.5$~$\mu$eV and
$\Delta_{m}=-72.5$~$\mu$eV. The other parameters are explained in
the text.}\label{fig:varrho-z}
\end{figure}

In Figs.~\ref{fig:varrho-eta}(a)-(f), we plot the conversion
efficiency $\rho$ as a function of $\eta$ with different propagation
distance (corresponding to different optical depth). The parameters
are chosen as $N=10^{23}$~m$^{-3}$,
$\gamma_{3}=\gamma_{4}\approx1.0$~meV, $\gamma_{5}\approx0.1$~meV,
$\Delta_{p}=0.5$~meV, $\Delta_{c}=3.0$~meV, $\Omega_{c}=2.0$~meV. As
discussed in Ref.~\cite{deng-prl-2002}, a phase-matched FWM field
can be produced if the reduced group velocity of the pump wave
$\omega_{p}$ subject to EIT is about equal to the group velocity of
the generated FWM wave $\omega_{m}$. The detuning of the generated
FWM field is therefore taken as $\Delta_{m}=72.5$~$\mu$eV
[Figs.~\ref{fig:varrho-eta}(a)-(c)] and $\Delta_{m}=-72.5$~$\mu$eV
[Figs.~\ref{fig:varrho-eta}(d)-(f)]. With these parameters, the
group velocities of the pump and FWM waves are matched
($v_{g}^{p}\approx v_{g}^{s}\approx3.7\times10^{-3}c$), which
enhances the field-medium interaction time and also results to an
increase of the FWM conversion efficiency of the system. For a
quantitative comparison with Ref.~\cite{deng-prl-2002} that treated
the EIT-enhanced FWM process, we have also plot the evolutions of
$\rho$ versus $\eta$ in the case of no resonant tunneling (i.e.,
taking $q=k=m=0$ such that the subband $|4\rangle$ is decoupled).
The result is shown by the dashed curves in
Figs.~\ref{fig:varrho-eta}(a)-(f). We see that, by virtue of
resonant tunneling, the conversion efficiency is enhanced
dramatically, especially for small propagation distance. For
examples, with $z=1.0$~$\mu$m and $z=10.0$~$\mu$m, the conversion
efficiency can be enhanced more than one order of magnitude around
the center frequency [see Figs.~\ref{fig:varrho-eta}(a), (b), (d),
and (e)]. Notice that, with increasing the propagation distance, the
behaviors of conversion efficiency with and without resonant
tunneling are totally different. Equation~(\ref{eq-1-wm}) indicates
that there are two generated waves with the same ultraslow group
velocity when phase matching is achieved. In the case of without
resonant tunneling, the interference between two waves leads to
oscillation [shown in Figs.~\ref{fig:varrho-eta}(c) and (f) with
dashed curves]. Direct comparison of these result implies that the
enhancement of FWM conversion efficiency can be achieved by
combining resonant tunneling-induced constructive interference in
Kerr nonlinearity and EIT. In the regime where the propagation
distance is small, the conversion efficiency of FWM can be enhanced
more than one order of magnitude than that without resonant
tunneling in the vicinity of center frequency.

In order to understand the influence of resonant tunneling on the
FWM process more clearly, we plot, in Fourier space, the FWM
conversion efficiency $\rho$ with different values of $\eta$ as a
function of propagation distance $z$ in
Figs.~\ref{fig:varrho-z}(a)-(f) for two set of parameters. The value
of $\Delta_{m}$ has been taken as $\Delta_{m}=72.5$~$\mu$eV and
$\Delta_{m}=-72.5$~$\mu$eV in the left and right columns. The other
parameters are same as those in Figs.~\ref{fig:varrho-eta}. In the
plot we took $\eta=\pm0.75$ which corresponds to the location of the
FWM peaks at $z=10.0$~$\mu$m without resonant tunneling (see
Figs.~\ref{fig:varrho-eta}). The FWM conversion efficiency with
resonant tunneling can be larger than that without resonant
tunneling, especially in the region near the center frequency. As
those illustrated in Refs.~\cite{sun-ol-2007,sun-oe-2012}, the
physical reason behind is the constructive interference to Kerr
nonlinearity induced by resonant tunneling [please see
Eq.~(\ref{eq-1-wm})]. As the propagation distance increases, owing
to resonant tunneling, $\rho$ increases rapidly, and the details in
very small propagation distance region have been shown in
Figs.~\ref{fig:varrho-z}(g) and (h). Deep inside the medium, the
effective interference limits the further production of the FWM
field. When the generated FWM wave becomes sufficiently intense,
excitation channels $|1\rangle\to|3\rangle$ and
$|1\rangle\to|4\rangle$ via $\Omega_{m}+2\Omega_{d}^{*}$ open and
become important. Consequently, this efficient backcoupling pathway
results to competitive multiphoton excitations of the FWM generating
state~\cite{wu-pra-2004}.

\begin{figure}
\includegraphics[width=6.0cm]{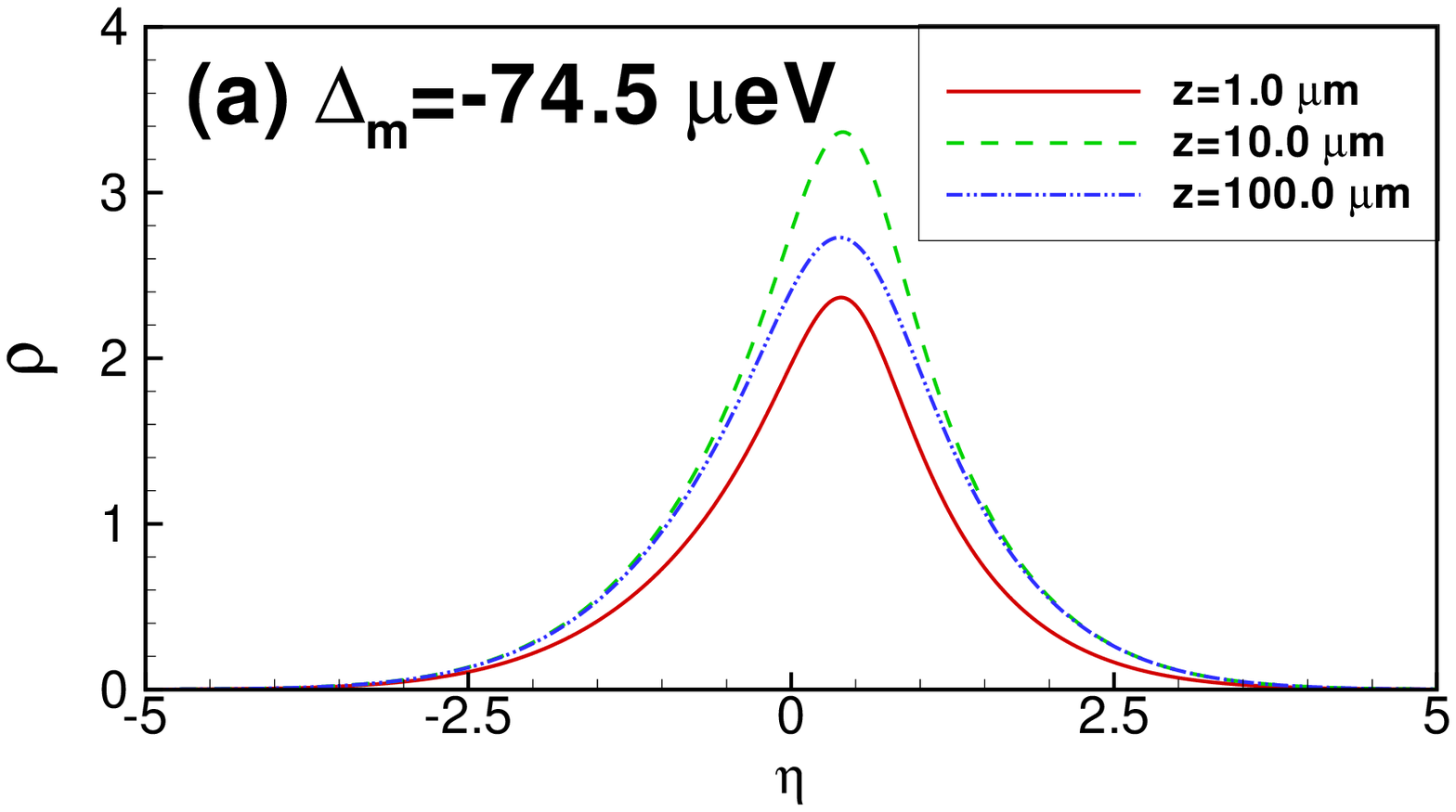}
\includegraphics[width=6.0cm]{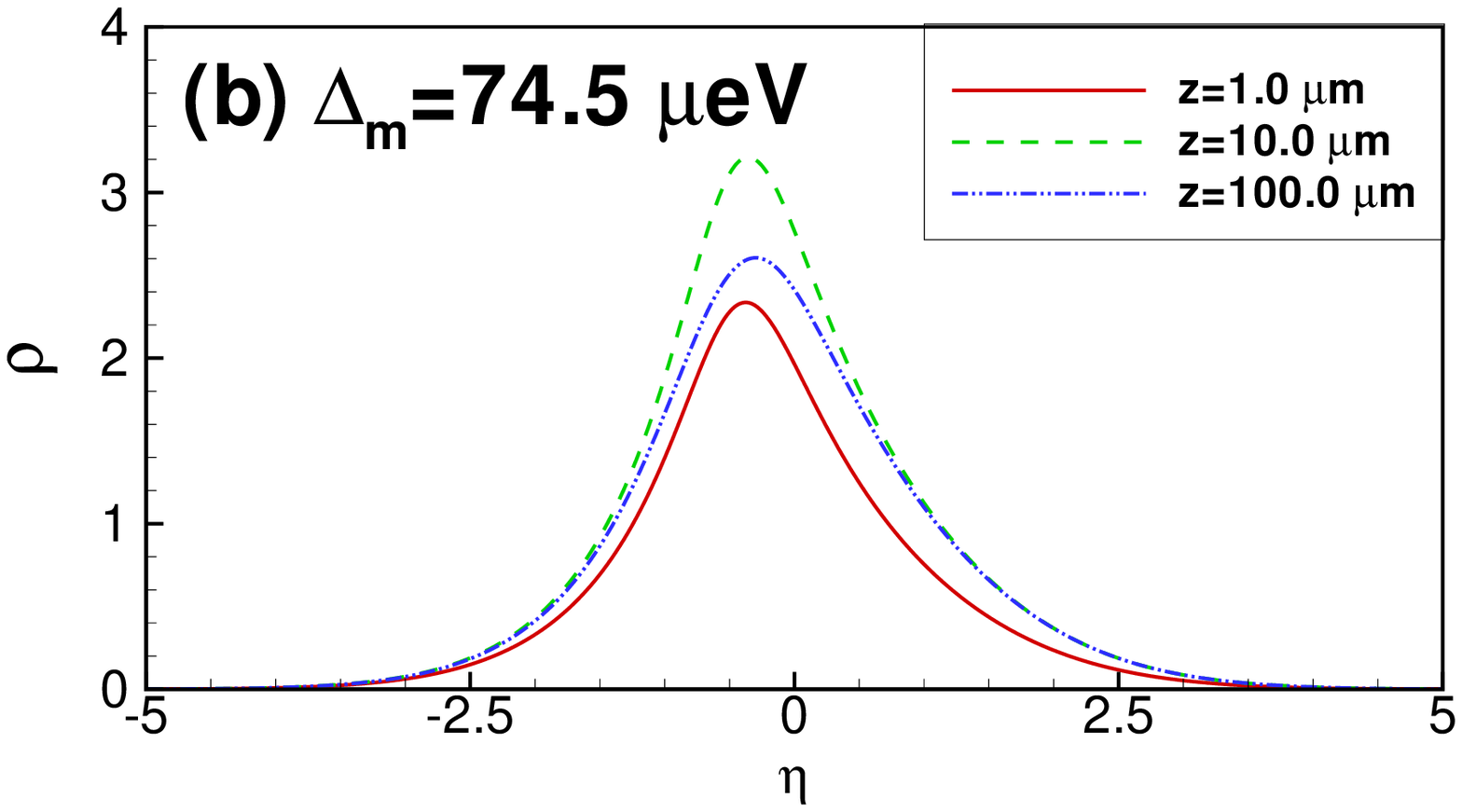}
\caption{(color online) The conversion efficiency of FWM $\rho$
versus $\eta$ without the control field
$\Omega_{c}$.}\label{fig:rho-eta-without-control}
\end{figure}

To demonstrate the role of the control field, we plot the FWM
conversion efficiency $\rho$ versus $\eta$ without the control field
in Figs.~\ref{fig:rho-eta-without-control}(a) and (b). At certain
pump detuning ($\Delta_{p}=4.0$~meV), the tunneling-induced
destructive interference between transition channels
$|3\rangle\to|1\rangle$ and $|4\rangle\to|1\rangle$ leads to the
negligible small absorption, i.e., TIT, which opens the channel for
FWM process. The physics behind is basically the same as EIT
illustrated in Refs.~\cite{deng-prl-2002,wu-pra-2004}. We choose the
detuning of the FWM wave as $\Delta_{m}=-74.5$~$\mu$m in
Fig.~\ref{fig:rho-eta-without-control}(a) and
$\Delta_{m}=74.5$~$\mu$m in
Fig.~\ref{fig:rho-eta-without-control}(b) such that the group
velocities of the pump and generated FWM waves are matched
($v_{g}^{p}\approx v_{g}^{m}\approx1.5\times10^{-2}c$). The results
with three different propagation distances are plotted. Direct
comparison of Figs.~\ref{fig:rho-eta-without-control} with the
corresponding results shown in Figs.~\ref{fig:varrho-eta}), we can
conclude that the presence of the control field modifies the linear
optical property of the QWs structure, as well as the FWM process.
As results, the matched reduced group velocity ensures larger
interaction time and also leads to an increase of the FWM conversion
efficiency.


In conclusion, we have suggested an asymmetric double QWs structure
to achieve parameter generation of a new laser radiation with high
conversion efficiency. This structure combines resonant tunneling
together with the advantages of EIT-enhanced FWM scheme. By virtue
of the constrictive interference in Kerr nonlinearity induced by
resonant tunneling, high efficiency of FWM process can be achieved.
Especially for small propagation distance and around the center
frequency, the FWM conversion efficiency can be enhanced more than
one order of magnitude than that without resonant tunneling.
Recalling the flexible design of QWs structure and the fact that the
resonant tunneling can be controlled by applying a bias voltage, we
believe that this excellent performance offers the possibility of
developing practical and portable electroptically modulated devices
based on FWM at low pump intensity.


We thank the financial support from the NSF-China under Grant No.
11104176, the NSF of Shaanxi Province under Grant No. 2011JQ1008,
the Fundamental Research Funds for the Central University under
Grant No. GK201003003, as well as the Open Fund from the SKLPS of
ECNU. S.L.F would like to acknowledge the support of NSF-China under
Grant No. 11104174.

\end{document}